\documentstyle[12pt,epsf]{article}
\def\be{\begin{equation}}
\def\ee{\end{equation}}
\def\bea{\begin{eqnarray}}
\def\eea{\end{eqnarray}}
\def\beaa{\begin{eqnarray} \begin{array}{c}}
\def\eeaa{\end{array} \end{eqnarray}}
\def\la{\lambda}

\def\ep{\epsilon}
\def\th{\theta}

\begin{document}
\begin{titlepage}
\setcounter{footnote}0
\begin{center}
\hfill Landau-97-TMP-1\\
\hfill ITEP-TH-5/97\\
\hfill hep-th/9702063\\
\vspace{0.3in}
\bigskip
\bigskip
{\bf Supersymmetric Spin Glass}\\
\bigskip
\bigskip
Sergei Gukov \footnote{E-mail
address: gukov@landau.ac.ru}\\
\bigskip
{\em L.D.Landau Institute for
Theoretical Physics
\\ 2, Kosygin st., 117334, Moscow, Russia\\
and\\
Institute of Theoretical and Experimental Physics\\
25, B.Cheremushkinskaya st., 117259, Moscow, Russia}\\
\bigskip
{\em 7 February 1997}\\
\end{center}
\begin{abstract}

The evidently supersymmetric four-dimensional
Wess-Zumino model with quenched disorder is
considered at the one-loop level. The infrared fixed
points of a beta-function form the moduli
space ${\cal M}=RP^2$ where two types of phases were
found: with and without replica symmetry. While the
former phase possesses only a trivial fixed point, this
point become unstable in the latter phase which may be
interpreted as a spin glass phase.

\end{abstract}

\end{titlepage}

\newpage

\setcounter{footnote}0

\begin{center}
{\bf \it The Devil Is Not So Black As He Is Painted.}
\end{center}

\bigskip

\section{Introduction}
\qquad
There is a great deal of field theory models
describing a system in quenched random fields
or coupling constants (\cite{Mezard},
\cite{Dotsenko}, \cite{CFT}, {\it etc}.). In solid state
physics such models naturally arise  the corresponding
pure systems whenever impurities are introduced.
It is interesting to extend randomness to other
well-studied field theories, just as, for example in
\cite{CFT}, disorder was implemented into minimal
conformal
models. As shown in \cite{Parisi} and subsequent papers
stochastic equations as well as the field theories in
presence of random external sources often prove to possess
some hidden supersymmetry. Kurchan \cite{Kurchan} indorsed
this result for spin glass dynamics. Because supersymmetry
handle perturbative corrections, such random theories are
especially interesting. This is what we will do in
this paper.

On the other hand in field theories with apparent
space-time supersymmetry superpotential is holomorphic
function not only of fields but also of coupling
constants \cite{Seiberg}.
Therefore couplings and fields enter potential
on equal footing, so that it seems very natural
to introduce random (gaussian) distribution of some
couplings in the Lagrangian. But the power of
supersymmetry is so strong that superpotential gets
no quantum corrections \cite{Seiberg}, \cite{Intr},
i.e. provided that the
coupling has no dynamical D-terms integrating it
over solves the problem.

In Section 2 we formulate four-dimensional supersymmetric
Wess-Zumino theory in random field.
In the context of replica method infrared fixed points of
one-loop $\beta$-functions are found in Section 3.
Analysis of these fixed points suggests two phases on the
moduli space ${\cal M}=RP^2$. Numerical evaluation of the
most general expressions is eventuated in the phase
diagram which is illustrated by two simple examples of
the fourth section. Section 5 is devoted to discussions
and conclusions.

\section{Wess-Zumino model perturbed by randomness}
\qquad
{}From above arguments it follows that the SUSY analog
of a theory with disorder must contain dynamical
terms for the random field. In the present paper
we consider a four-dimensional Wess-Zumino model that
is the supersymmetric counterpart of $\varphi^4$-model
(the both theories are defined in the same critical
dimension
and the scalar potential after integrating auxiliary
field in the former model is actually $\varphi^4$).
Since, according to \cite{Intr}, Wess-Zumino theory is
defined only as a low-energy field theory, we will study
the Wilsonian effective action by integrating fast modes
with momentum $\Lambda'<P< \Lambda$.
Thereby, let us
define a chiral superfield $\Phi=\varphi + \th \psi
+ \th^2 F$ and a random superfield $H$. In this
notations the original action is\footnote{For the sake
of simplicity the mass terms are omitted.}:
\bea
S=\int d^4x d^2\th d^2 \bar \th (g \Phi^{+} \Phi -
\Phi^{+}H-H^{+}\Phi+{1 \over u}H^{+}H) +\nonumber \\
+ {1 \over 3!} \int d^4x d^2\th ( \la_1' \Phi H^2 +
\la_2' \Phi^2 H + \la_3' \Phi^3 + \la_4' H^3) + h.c.
\label{action}
\eea

This action admits the following treatment. It may
be obtained (for the certain set of parameters) from
the usual Wess-Zumino action by the replacement
$\Phi \rightarrow \Phi + H$, as one usually does
in summation over local extremes \cite{Dotsenko}.

One of the most powerful method to deal with random
fields is the replica trick \cite{Mezard}, which we
will use here to solve this
"toy" model. It reduces to introducing $n$ copies
(replicas) of our system, integrating $H$ field out,
then solving $n$-replica problem and taking $n=0$ at
the end of calculations. After replication the action
(\ref{action}) takes the form :
\bea
S=\int d^4x d^2\th d^2 \bar \th [ \sum_{a=1}^n
(g \Phi_a^{+} \Phi_a - \Phi_a^{+}H - H^{+}\Phi_a)+
{1 \over u}H^{+}H] +\nonumber \\
+{1 \over 3!}
\int d^4x d^2\th [ \sum_{a=1}^n
(\la_1' \Phi_a H^2 + \la_2' \Phi_a^2 H +
\la_3' \Phi_a^3) + \la_4' H^3] + h.c.
\label{replact}
\eea

As will be shown later the
model depends only on the relative values of lambdas,
so that one can put them small enough to determine
$H$ field from the saddle point equation on D-term only:
\be
H=u \sum_{a=1}^n \Phi_a \qquad and \qquad
H^{+}=u \sum_{a=1}^n \Phi_a^{+}
\label{h}
\ee

Substituting it back into (\ref{replact}) yields:
\beaa
S= \sum_{a,b=1}^n \int d^4x d^2\th d^2 \bar \th g_{ab}
\Phi_a^{+} \Phi_b  + \label{act} \\
+{1 \over 3!} \int d^4x d^2\th
( \sum_{a,b,c=1}^n \la_1 \Phi_a \Phi_b \Phi_c +
\sum_{a,b=1}^n \la_2 \Phi_a^2 \Phi_b +
\sum_{a=1}^n \la_3  \Phi_a^3 ) + h.c.
\eeaa
where
$g_{aa}=g+3u$, $g_{a \ne b}=3u$
and
three types of vertexes
$\la_1=\la_1' u^2+\la_4' u^3$, $\la_2=\la_2' u$,
$\la_3=\la_3'$
band differently replica indices.
It is the action (\ref{act}) that we are going to study.

\section{Fixed points of $\beta$-functions}
\qquad
Renormalisation group (RG)
equations for $g_{ab}$ easily follow from the one-loop
diagram for the pure Wess-Zumino theory \cite{Wess}:
\bea
{dg_{ab} \over d \ln{\Lambda}} = {1 \over 288 \pi^2}
\{9\la_3^2 g_{ab}^2 + 2\la_2^2\sum_{c,d=1}^n [(g_{ac}+
g_{bc})g_{cd}+g_{ac}g_{bd}]+ \nonumber \\
+3\la_2 \la_3\left[ \sum_{c=1}^n (g_{ac}^2+g_{bc}^2)
+ 2 g_{ab} \sum_{c=1}^n (g_{ac} + g_{bc}) \right] +
9 \la_1 \la_3 \sum_{c,d=1}^n (g_{ac}g_{ad}+g_{bc}g_{bd})\}
\label{beta}
\eea

Taking into account the possible replica symmetry breaking
we take the Parisi ansatz for $g_{ab}$ \cite{Mezard}:
off-diagonal part of $g_{ab}$ is parametrized by internal
function $g(x)$ defined on a unite interval
$x \in [ 0,1 ]$ and diagonal part is $g_{aa} = \tilde g$.
Replica-symmetric case is obtained by putting
$g(x)=g=constant$.  Algebra of Parisi matrices ${\bf a}=
(\tilde a, a(x))$ is defined by the multiplication rule
\cite{Mezard}:
\bea
{\bf c}={\bf ab}: \qquad \tilde c= \tilde a \tilde b -
\int^1_0 dx a(x)b(x) \nonumber \\ c(x)=b(x) [\tilde a -
\int^1_0 dx a(y) ] + a(x) [\tilde b -
\int^1_0 dx b(y) ] -\label{mult} \\
-\int^x_0 dy (a(x)-a(y))(b(x)-b(y))
\nonumber
\eea

By means of this rule we get sums over replica indices
that appear in (\ref{beta})
in the $n \rightarrow 0$ limit:
\be
\sum_{b=1}^n g_{ac} \rightarrow \tilde g - \bar g \qquad
\sum_{c,d=1}^n g_{ac} g_{cd} \rightarrow (\tilde g -
\bar g)^2
\qquad \sum_{b=1}^n g^2_{ac} \rightarrow \tilde g^2 -
\bar{g^2}
\nonumber
\ee
where
\be
\bar g = \int^1_0 dx g(x) \qquad and \qquad
\bar{ g^2} = \int^1_0 dx g^2(x)
\nonumber
\ee

A question arises, as usual in spin glass theory, to find
infrared (IR) fixed points\footnote{The points where
$\beta$-functions vanish.} of (\ref{beta}) which
determine the dynamics of the system:
\bea
{3 \over 2}\la_3^2 \tilde g^2 + (\la_2^2+3\la_1 \la_3)
(\tilde g - \bar g)^2 + \la_2 \la_3 [2 \tilde g (\tilde g
- \bar g) + \tilde g^2 - \bar{g^2}]=0
\label{rg} \\
{3 \over 2}\la_3^2 g^2(x) + (\la_2^2+3\la_1 \la_3)
(\tilde g - \bar g)^2 + \la_2 \la_3 [2 g(x) (\tilde g
- \bar g) + \tilde g^2 - \bar{g^2}]=0
\label{rgg}
\eea

For example, the $\la_2^2$-term is produced by the two
non-vanishing (with the number of replicas) diagrams shown
on Fig.1.
\begin{figure}
\epsfxsize 400pt
\epsffile{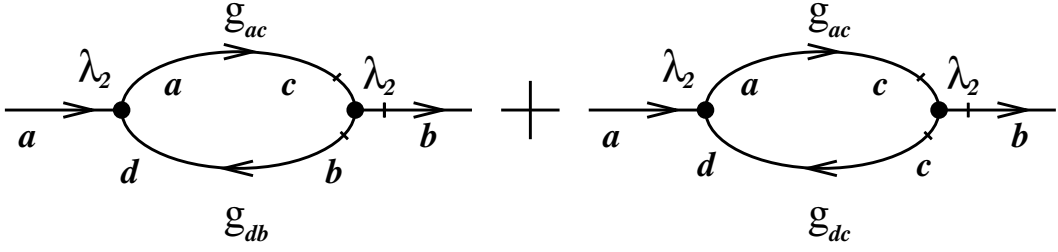}
\caption{Surviving (in the $n \rightarrow 0$ limit)
$\la_2^2$-contributions.}
\end{figure}

These equations have
two remarkable properties: they are homogeneous in $\la$
and $g$, i.e. depend only on the squares of the both.
Such dependence on $\la$ tells us that zeroes of
beta-functions  (\ref{rg})-(\ref{rgg}) do not depend
on the values of the couplings themselves,
but only on their mutual ratios, so that the moduli
space of the theory is
$RP^2$ instead of $R^3= \{ \la_1,\la_2,\la_3 \}$.
Therefore, without loss of
generality, we may put couplings very small keeping
their ratios fixed. In this limit the results that
we are going to obtain are exact.
Moreover, in what
follows we will assume $\la_3 \ne 0$, so
that we can choose it to be $\la_3=1$ and denote
$\la_2=\la$ and $\la_1=\mu$ (affined map). \footnote{If
$\la_3 \ne 1$ then the right
parameters are $\la = {\la_2 \over \la_3}$ and $\mu =
{\la_1 \over \la_3}$.} The special case of $\la_3=0$
will be studied in the first example of Section 4.

Quadratic dependence on $g$ in (\ref{rgg})
means that for each set of general characteristics, such
as $\bar g$, $\bar{g^2}$ and $\tilde g$, there are only
two possible values $g_{1,2}$ (if any) which the function
$g(x)$ can take in a IR-fixed point. Moreover, the same
must be true for $\tilde g$ because formally it also
satisfies similar equation (\ref{rg}). We are free
to chose
$\tilde g = g_1$, for instance. Let us denote the measure
of points
on a unite interval of $x$ where $g(x)=g_1$ as $1-x_0$ and
the measure
of points where $g(x)=g_2$ as $x_0$. For example, it may
be a stepwise distribution:
\begin{eqnarray}
g(x)=\left\{
\begin{array}{ccc}
g_1, \qquad x_0<x<1
\\
g_2, \qquad 0<x<x_0
\label{param}
\end{array}
\right.
\end{eqnarray}

Thus we have two equations (\ref{rg})-(\ref{rgg}) on
three quantities $g_{1,2}$ and $x_0$ with
$\bar g$ and $\bar{g^2}$ depending on them. If $g_1$
and $g_2$ are not simultaneously equal to
zero\footnote{Otherwise we get a trivial
replica-symmetric fixed point.} then, actually, we have
only two unknowns: $x_0$ and the
ratio $p={g_2 \over g_1}$.
In this notations (\ref{rg})-(\ref{rgg}) may be rewritten
as:
\begin{eqnarray}
\left\{
\begin{array}{ccc}
1+({2\over3} \la^2 +2 \mu) x_0^2 (1-p)^2 +
{2 \over 3} x_0
\la \left[ 2(1-p) + (1-p^2) \right] = 0 \label{last} \\
p^2+({2\over3} \la^2 +2 \mu) x_0^2 (1-p)^2 +
{2 \over 3} x_0
\la \left[ 2p(1-p) + (1-p^2) \right] = 0
\end{array}
\right.
\end{eqnarray}
determining both $p$ and $x_0$, and, consequently,
the phase of the system.

Curiously enough, for a given solution $p$ and $x_0$
we get the whole set of RG-fixed points $\{ \tilde g,
g(x)\}$ differing in arbitrary factor. Of course,
this degeneracy will be removed by higher loop
corrections,
so that particular value of the fixed point will be
determined by the full perturbative expansion.
At the one-loop approximation, the explicit data
$(\tilde g, g(x))$ in the fixed point may be determined by
the initial conditions $g$ and $u$.

If for some set of couplings there
is no solution to (\ref{last}) except the trivial
one $\tilde g= g(x)=0$, we will refer to this point
on the phase space $\{ \la, \mu \} \in {\cal M}=RP^2$
as a replica-symmetric point and will denote the
corresponding phase "RS". Otherwise, replica symmetry
is broken with $x_0$ being the solution of (\ref{last}),
and the corresponding phase "RSB" looks like a
spin glass system.

Since (\ref{last}) must be solved by the same $p$,
equating the solutions to each equation we get the relation
between $x_0$ and $\{ \la, \mu\} \in {\cal M}$.
Instead of writing the resulting complicated formulae
(partly because it can not be resolved relatively $x_0$),
we display it for $x_0=1$:
\be
\frac{\la^2+3\mu+\la \pm \sqrt{{5 \over 2}\la^2 -
{9 \over 2} \mu + {3 \over 2} \la}}{\la^2 +3 \mu -\la} =
\frac{\la^2+3\mu-\la \pm \sqrt{{5 \over 2}\la^2 -
{9 \over 2} \mu -{3 \over 2} \la}}{{3\over 2}+\la^2
+3 \mu - 3 \la}
\label{gen}
\ee
where sings in the both parts are taken independently.
Replacing $\la \rightarrow x_0 \la$ and
$\mu \rightarrow x_0^2 \mu$, we
get the equation (\ref{gen}) for arbitrary $x_0$.
This expression describes (part of) a curve in ${\cal M}$
that separates RS and RSB phases as shown on Fig.2.
The shaded region indicate replica-symmetric phase and
the unshaded region corresponds to replica symmetry
breaking where there is a non-trivial solution to
(\ref{gen}), and the trivial point $\tilde g = g(x) =0$
becomes unstable,
as will be descanted in the second example of the
next section.

\begin{figure}
\epsfxsize 400pt
\epsffile{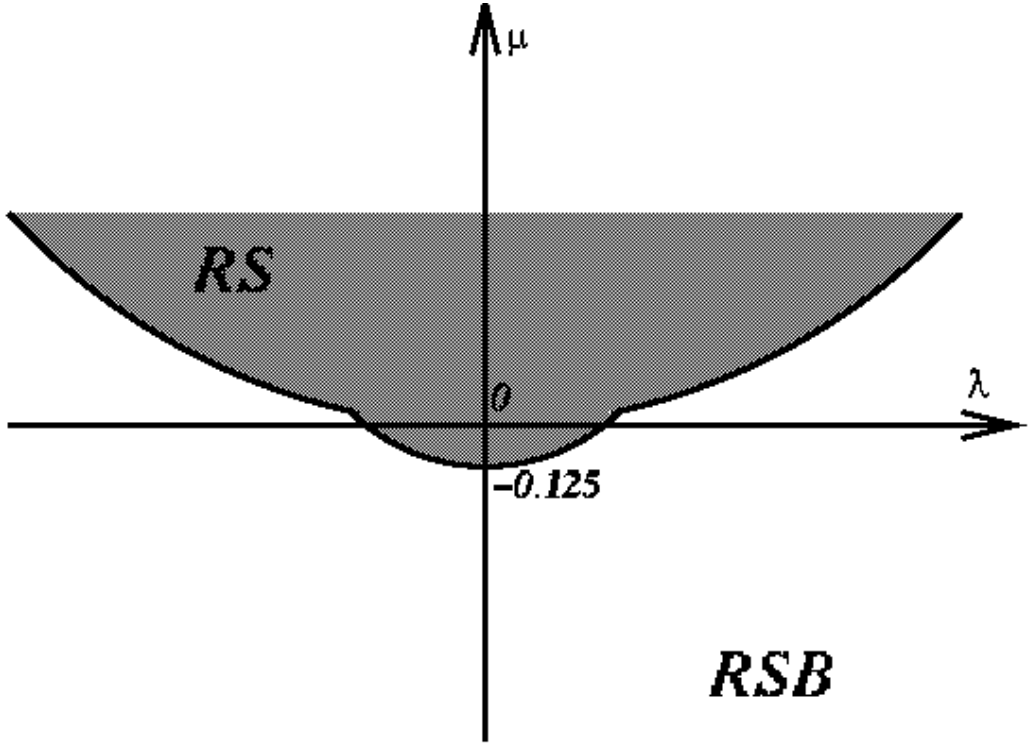}
\caption{The Phase Diagram (not drawn in scale).}
\end{figure}

\section{Two simple examples}

\begin{itemize}
\item $\la_3=0$ \\
In this case the beta-functions (\ref{beta}) become
\bea
{d \tilde g \over d \ln{\Lambda}} = {1 \over 48 \pi^2}
\la_2^2(\tilde g - \bar g)^2 \nonumber \\
{d g(x) \over d \ln{\Lambda}} = {1 \over 48 \pi^2}
\la_2^2(\tilde g - \bar g)^2 \nonumber
\eea

These equations may be easily integrated with the result:
\bea
\tilde g_{\Lambda} = \tilde g_{ 0,\Lambda' }
+ {A \over 48 \pi^2}\la_2^2
\ln{{\Lambda \over \Lambda' }} \nonumber \\
g_{\Lambda} (x) = g_{ 0, \Lambda' } (x)
+ {A \over 48 \pi^2} \la_2^2
\ln{{\Lambda \over \Lambda'}}
\nonumber
\eea
where a constant $A=(\tilde g - \bar g)^2$ is determined
by initial conditions and remains unchanged during
renormalisation group flow. Since for any $\la_2$ the only
fixed point is $\tilde g = g(x)=0$, this phase is
always replica-symmetric and is not as
interesting as others. \\

\item $\la_2=0 \leftrightarrow \la=0$ \\
Equations (\ref{rg})-(\ref{rgg}) take the form:
\begin{eqnarray}
\left\{
\begin{array}{ccc}
\tilde g^2 + 2 \mu (\tilde g - \bar g)^2 =0
\label{r20}
\\
g^2 (x) + 2 \mu (\tilde g - \bar g)^2 =0
\label{r201}
\end{array}
\right.
\end{eqnarray}

for which $g_{1,2}= \pm g$ for some $g \ne 0$ in
the SG phase. In parametrization (\ref{param})
\be
\bar g = (g_1-g_2)x_0 = 2gx_0 \qquad and \qquad
\bar g^2 = (g_1^2-g_2^2)x_0 = 2g^2 x_0
\ee

Substitution it into (\ref{r20}) yields a nontrivial
solution:
\be
-8 \mu x_0^2 =1 \qquad or \qquad x_0 =
{1 \over \sqrt{-8 \mu}}
\label{s20}
\ee
which exists only for $\mu < - {1 \over 8}$. It is the
range of $\mu$ where the RSB phase can be found.
Let us emphasize that exactly for these points
in ${\cal M}$ the trivial fixed
point $\tilde g = g(x)=0$ becomes unstable, for example,
with respect to perturbations in $\tilde g$.
To see this consider $\tilde g= \ep$:
\be
{d \ep \over d \ln{\Lambda}} = \alpha \ep^2
\nonumber \\
\ee
where $\alpha<0$ if (\ref{s20}) is true (i.e. arbitrary
small $\ep$ increases during the flow to low energies).
This simple case illustrates the behavior of general system
(\ref{last}). On the phase diagram it corresponds to
$\mu$ axis where the both RS and RSB phases exist.
\end{itemize}

\section{Summary}
\qquad
Having started from the (space-time)
supersymmetric Wess-Zumino model in a random and quenched
background (\ref{action}) we have found that
renormalisation group equations (\ref{beta}) in a fixed
point are quadratic
homogenous equations in couplings and in $g$. The former
property allowed us to take couplings very small as well
as to reduce the moduli space to ${\cal M}=RP^2$. There
are two types of points (phases) on this moduli space with
either broken replica symmetry or not.

Though we have found all IR-fixed points of one-loop
$\beta$-function, stability of nontrivial fixed points
and analytic RG flow to them remain unexplored.
Finally, it is interesting to generalize this analysis
to more complex supersymmetric theories as well as to find
realistic models whose critical behavior correspond
to such theories.

\vskip5mm
\section*{Acknowledgments}
\qquad
I would like to thank D.Ivanov, A.Marshakov and A. Mironov
for wholehearted atmosphere and helpful conversations.
I am especially indebted to A.Morozov and I.Polyubin for
their stimulating suggestions and comments. Also I am
grateful to Vik. Dotsenko from whom I had for the first
time known what spin glass is, and with whom many of the
ideas presented here have been discussed.

This work was supported in part
by RFBR grant No. 96-15-96939.

\end{document}